%% file: main.tex
\documentclass{sig-alternate-10pt}

\usepackage{pslatex}
\usepackage{balance}
\usepackage{xcolor}
\usepackage[margin=1in]{geometry}

\usepackage{hyperref}
\usepackage{authblk}
\usepackage{enumitem}



\begin{document}
\pagestyle{empty}

\renewcommand\Authfont{\fontsize{11}{14.4}\selectfont}
\renewcommand\Affilfont{\fontsize{9}{10.8}}

\newcommand{\MD}[1]{\textcolor{blue}{#1}}

\title{LLM+Graph@VLDB'2025 Workshop Summary}

% \author{Yixiang Fang$^1$\,\, Arijit Khan$^2$\,\, Tianxing Wu$^3$\,\, Da Yan$^4$\,\, Shu Wang$^1$
% \\
% \affaddr{$^1$The Chinese University of Hong Kong, Shenzhen\,\, $^2$Aalborg University\,\, $^3$Southeast University}
% \\ 
% \affaddr{$^4$Indiana University Bloomington}
% \\
% \email{\small fangyixiang@cuhk.edu.cn\,\,arijitk@cs.aau.dk\,\,tianxingwu@seu.edu.cn\,\,yanda@iu.edu\,\,shuwang3@link.cuhk.edu.cn}
% }

\author[1]{Yixiang Fang}
\author[2]{Arijit Khan}
\author[3]{Tianxing Wu}
\author[4]{Da Yan}
\author[1]{Shu Wang}

\affil[ ]{\textsuperscript{1}The Chinese University of Hong Kong, Shenzhen, \textsuperscript{2}Bowling Green State University, \textsuperscript{3}Southeast University, \textsuperscript{4}Indiana University Bloomington}

\affil[ ]{\textsuperscript{1}\{\texttt{fangyixiang@, shuwang3@link.\}cuhk.edu.cn}, \textsuperscript{2}\texttt{arijitk@bgsu.edu}, \textsuperscript{3}\texttt{tianxingwu@seu.edu.cn},\textsuperscript{4}\texttt{yanda@iu.edu}
}

\maketitle

\input{abstract}

\input{1-INTORDUCTION}
\input{2-KEYNOTES}
\input{3-INDUSTRIAL}
\input{4-RESEARCH}
\input{5-PANEL}
\input{6-Future}

\clearpage
\balance

\bibliographystyle{abbrv}
\bibliography{refs}
\vspace{6mm}

\end{document}

%% file: abstract.tex
\begin{abstract}
%The synergy between large language models (LLMs) and graph-structured data has emerged as a crucial and rapidly developing research area, attracting significant attention from both academia and industry.
%
%At the 2nd workshop, LLM+Graph, co-located with 51$^{st}$ International Conference on Very Large Data Bases (VLDB) 2025 in London, the key theme explored was the inspection of effective algorithms and systems to unite LLMs, graph data management, and graph machine learning for real-world applications. This report outlines the major research directions, challenges, and innovative approaches presented by the various speakers during the workshop.
The integration of large language models (LLMs) with graph-structured data has become a pivotal and fast evolving research frontier, drawing strong interest from both academia and industry. The 2nd LLM+Graph Workshop, co-located with the 51st International Conference on Very Large Data Bases (VLDB 2025) in London, focused on advancing algorithms and systems that bridge LLMs, graph data management, and graph machine learning for practical applications. This report highlights the key research directions, challenges, and innovative solutions presented by the workshop's speakers.

\end{abstract}

%% file: 1-INTORDUCTION.tex
\section{Introduction}

\medskip
\medskip

%Large Language Models (LLMs) such as ChatGPT and LLaMA have seen rapid development, revolutionizing natural language processing with applications across numerous domains. While LLMs are proficient at learning from massive text corpora, they often lack consistent, factual knowledge representations, leading to unreliable responses and hallucinations. In parallel, graph-structured data is ubiquitous, modeling complex relationships in everything from social networks to knowledge bases. The synergy between the parametric knowledge of LLMs and the explicit, structured knowledge of graphs has therefore become a critical research frontier. This synergy is bi-directional: on one hand, graphs can enhance LLMs through techniques like graph-based Retrieval-Augmented Generation (GraphRAG) \cite{zhou2025depth,wang2025archrag} to provide contextual facts and improve accuracy. On the other hand, LLMs can facilitate downstream graph tasks such as knowledge graph (KG) construction, graph foundation models, graph data management and mining \cite{}.

Large Language Models (LLMs) such as ChatGPT and LLaMA have advanced rapidly--transforming natural language processing with broad applications across domains. Despite their ability to learn from massive text corpora, LLMs often lack stable factual representations, resulting in unreliable outputs and hallucinations. In contrast, graph-structured data is pervasive, capturing complex relationships in domains ranging from social networks to knowledge bases. The interplay between the parametric knowledge of LLMs and the explicit, structured knowledge of graphs has thus emerged as a critical research frontier. This synergy is bidirectional: graphs can strengthen LLMs through methods such as graph- and knowledge graph-based Retrieval-Augmented Generation (GraphRAG and KG-RAG) \cite{abs-2501-08686,zhou2025depth,wang2025archrag}, improving contextual accuracy, while LLMs can advance graph-centric tasks including knowledge graph construction~\cite{zhang2024extract}, graph foundation models~\cite{sun2025riemanngfm}, graph data management and mining~\cite{chat2graph,terdalkar2025graph,pachera2025user}.
% \textcolor{red}{\cite{}}.

%The ambition of the 2nd workshop on data management opportunities in bringing LLMs with graph data (LLM+Graph@VLDB'2025) was to provide a unique platform for researchers and practitioners to present the latest research, new technologies, and emerging applications in this trending area. Building on the success of the previous year's workshop on LLM+KG, this edition expanded its scope beyond KGs to cover the broader fields of graph computing, data management, and machine learning, aiming to draw attention to the exciting opportunities awaiting data management researchers in this greener pasture.

The 2nd Workshop on Data Management Opportunities in Integrating LLMs with Graph Data (LLM+Graph @VLDB25) aimed to provide a dedicated forum for researchers and practitioners to showcase cutting-edge research, emerging technologies, and novel applications in this rapidly evolving field. Building on the success of the inaugural LLM+KG workshop@VLDB24 \cite{KhanWC25}, this edition broadened its scope beyond knowledge graphs to encompass graph computing, data management, and learning, highlighting the expanding opportunities for data management researchers in this dynamic area.

The full-day workshop had 3 keynote talks on the union of LLMs, graph data management, and graph learning, 3 industrial invited talks on GraphRAG applications, LLMs on Neo4j graph database \cite{llmneo4j}, and Chat2Graph \cite{chat2graph}, 7 peer-reviewed research papers at the intersection of LLMs and graph technologies, and a panel discussion on the topic of ``LLMs vs. Graphs: Supercharge or Supersede Graph Data Management, Mining, and Learning?'' The detailed program is available at \cite{fang2150llm+}.

%% file: 2-KEYNOTES.tex
\section{keynotes}

\medskip
\medskip

The program featured three keynotes by Prof. M. Tamer Özsu (University of Waterloo), Prof. Philip S. Yu (University of Illinois, Chicago), and Prof. Angela Bonifati (Lyon 1 University).

\subsection{The Duality Between Large Language Models and Database Systems}

\medskip
\medskip
The first keynote was delivered by Prof. M. Tamer Özsu from the University of Waterloo, who addressed the duality between LLMs and database systems. He argued that while LLMs challenge the core mission of data management, there is a symbiotic relationship: LLMs can assist database tasks (LLM4DB), and databases can empower LLMs (DB4LLM)

In the \textbf{LLM4DB} direction, Prof. Özsu explored how LLMs are being leveraged to enhance data management. Key research areas include Natural Language Querying, where he presented an interactive approach that uses a natural language explanation of a preliminary query as a ``handle'' for users to provide feedback~\cite{jian2025interacsparql}. 
% \textcolor{red}{\cite{}}. 
He also addressed the semantic evaluation of NL2SQL models, highlighting the limitations of current metrics and introducing SQLyzr, a comprehensive benchmark for more granular, multi-faceted evaluation~\cite{abedini2026sqlyzr}.
% \textcolor{red}{\cite{}}.
A third area covered was efficient LLM inference over existing data by embedding LLM capabilities as user-defined functions, noting that systems like FLOCKMTL \cite{dorbani2025beyond} and PGAI \cite{akillioglu2150research} are already exploring this space.

Conversely, in the DB4LLM direction, Prof. Özsu highlighted how data management principles are critical for empowering LLMs. A major focus was on approximate nearest neighbor search in vector databases and indexing. He contrasted the trade-offs between refinement-based methods (e.g., KGraph \cite{dong2011efficient}, NNDescent \cite{bratic2018nn}) and increment-based methods (e.g., HNSW \cite{malkov2018efficient}). He then introduced his group’s solution, MIRAGE \cite{voruganti2025mirage}, a novel graph-based indexing approach that achieves both high indexing speed and superior search performance. Finally, he discussed RAG over multimodal databases, emphasizing that the next step is enabling complex reasoning over retrieved heterogeneous data.

Prof. Özsu concluded by acknowledging the potential of LLMs but cautioned against current limitations like hallucination and high cost~\cite{zhou2024llm}. He reinforced the symbiotic relationship between the fields and the vital role of the data management community in building the underlying technology.

\subsection{Towards Graph Foundation Models \\ with Riemannian Geometry}

\medskip
\medskip
The second keynote was delivered by Prof. Philip S. Yu from the University of Illinois, Chicago, who addressed the challenge of developing a true graph foundation model.
He argued that the current Euclidean GNN paradigm is insufficient for representing complex graphs and introduced Riemannian Geometric Deep Graph Learning (RiemannGeoDGL)~\cite{sun2025riemanngfm} as a more expressive alternative that can better model non-Euclidean structures like hierarchies and cycles.

The talk detailed several key research directions within this paradigm. Prof. Yu introduced the use of heterogeneous (mixed) curvature spaces, often product spaces of hyperbolic and spherical manifolds, to capture the diverse geometries within a single complex graph.
This approach provides different geometric ``views'' for powerful self-supervised learning. 
He also explained how geometric concepts like Ricci curvature can be leveraged for tasks such as graph clustering, leading to methods like Congregate \cite{sun2023congregate} for contrastive clustering and RicciNet \cite{sun2024riccinet} for clustering irregularly shaped data.
Furthermore, he showed the framework's applicability to graph dynamics, including modeling physics-informed systems with Pioneer \cite{sun2025pioneer} and transferable source localization with Trace \cite{suntrace}.

Ultimately, these advancements converge on the goal of creating a Riemann Graph Foundation Model (RiemannGFM) \cite{sun2025riemanngfm}. Unlike traditional methods, RiemannGFM focuses on graph structure by proposing a universal structural vocabulary of trees and cycles. By mapping this vocabulary to its corresponding Riemannian geometry, the model can learn generalizable structural representations. 
Prof. Yu also noted that this geometric approach offers a new way to design deeper models that can overcome common issues like oversmoothing and oversquashing.

In conclusion, Prof. Yu posited that RiemannGeoDGL opens a promising new direction for building true graph foundation models that are structurally aware, highly generalizable, and capable of greater depth by embracing the inherent geometry of complex networks.

\subsection{Reasoning over Property Graphs: Le-veraging Large Language Models for Automated Data Consistency}
 
\medskip
\medskip
The third keynote, by Prof. Angela Bonifati from Lyon 1 University, addressed the challenge of repairing semantic inconsistencies in property graphs, first via a user-centric framework and then by exploring the capabilities of LLMs.
 
Prof. Bonifati first argued that ambiguity in automated graph repair necessitates human intervention. She introduced cQAR \cite{pachera2025user}, a user-centric framework to collaboratively involve users in the repair process, and highlighted a key theoretical finding: ensuring the safety of repairs is a necessary and sufficient condition for convergence
 
The talk then pivoted to \textit{whether LLMs can correct graph inconsistencies}. Different LLM tools and different graph-to-text encodings have been tested and cross-compared in an extensive empirical study \cite{terdalkar2025graph}. Prof. Bonifati concluded that LLMs are not yet reliable for this task, as the correctness of suggested repairs was very low despite often having a valid format. Common failure modes included eager generation, indecision, and hallucination.
 
Finally, Prof. Bonifati explored the inverse problem of \textit{using LLMs for graph rule generation} \cite{LeB025}. Existing approaches would generate several redundant graph rules and rules that are not meaningful in practice. She presented a pipeline that uses a sliding window attention mechanism and RAG to manage the graph's size before prompting an LLM to generate consistency rules and their corresponding Cypher queries.
 
In conclusion, Prof. Bonifati stated that LLMs are not yet ready for unsupervised graph repair, emphasizing the need for interactive, human-in-the-loop systems and better explainability to effectively leverage their potential for ensuring data consistency on graphs \cite{pan2023large}.

%% file: 3-INDUSTRIAL.tex
\section{Industrial invited talk}

\medskip
\medskip
The workshop program was also enriched by three industrial invited talks from Dr. Cheng Chen (ByteDance), Dr. Brian Shi (Neo4j), and Dr. Heng Lin (Ant Group).

\subsection{Applications and Challenges of Graph-RAG and Graph Foundation Models at ByteDance}

\medskip
\medskip
The first industrial invited talk was delivered by Dr. Cheng Chen from ByteDance. Cheng began by describing the massive scale of graph data at ByteDance, managed by their ByteGraph system, which powers applications like Douyin and Toutiao~\cite{li2022bytegraph}.

The first part of the talk focused on the practical applications of GraphRAG.
Cheng detailed three primary use cases: Code Understanding \& Generation, where GraphRAG helps capture dependencies across code modules~\cite{liu2024marscode}; automated Test Case Generation for new user interface (UI) features by organizing business logic into a graph~\cite{kong2025prophetagent}; and high-stakes QA for domains like legal compliance~\cite{chung2025graphcompliance}, where its ability to perform multi-hop reasoning and filter noise is essential. He also touched upon their modular GraphRAG architecture~\cite{li2022bytegraph}, which allows for flexible composition of different components.

Next, Cheng transitioned to their work on a Graph Foundation Model, motivated by the challenges of current methods. He focused on a key application in fraud detection, explaining the limitations of existing prompting strategies: encoding-based methods lose semantic richness, while text-only methods suffer from context domination. To solve this, his team developed Dual-Granularity Prompting ~\cite{li2025dgp}. 
This approach balances detail and efficiency by preserving fine-grained information for the target node while summarizing the context from its metapath-based neighbors, allowing the LLM to make more accurate predictions.

In conclusion, Cheng noted that GraphRAG is already widely deployed in over 50 scenarios at ByteDance. He stated that their long-term vision is to develop a unified Graph Foundation Model that can be trained once and applied across diverse business domains.

\subsection{Retrieval and Reasoning with LLMs on Neo4j: Progress and Challenges}

\medskip
\medskip
The second industrial invited talk was presented by Brian Shi from Neo4j.
Brian presented a comprehensive overview of Neo4j's efforts to integrate LLMs with their native graph database, showcasing a suite of tools designed to make graph data more accessible. He covered three main areas of innovation: GraphRAG, Text-to-Cypher, and a Graph Data Science (GDS) Agent.

The talk first covered their \textit{GraphRAG} framework, which employs a pipeline of vector retrieval, graph traversal, and pruning to generate a rich subgraph context for a GNN+LLM component~\cite{neo4j2025graphrag}.
% \textcolor{red}{\cite{}}. 
Brian presented results showing that this framework significantly improves retrieval accuracy when compared against baselines, including G-Retriever~\cite{he2024g} and GraphRAFT \cite{clemedtson2025graphraft}. He also noted key challenges, including neighborhood explosion in dense graphs and correctly identifying answer nodes that are many hops away from the initial entities.

Next, Brian introduced the \textit{Text-to-Cypher}~\cite{ozsoy2025text2cypher} project, highlighting Neo4j's contributions to the community, including public datasets like Text2Cypher-2025v1~\cite{Text2Cypher2025v1} and several fine-tuned LLMs~\cite{Text2CypherModels}. 
He also detailed their work on cost and performance optimization, presenting techniques such as Schema Filtering~\cite{ozsoy2025enhancing} to reduce inference tokens. 
Finally, Brian presented the \textit{GDS Agent} \cite{shi2025gds}, a reasoning agent that addresses complex analytical questions by autonomously selecting, parameterizing, and executing appropriate algorithms from the Neo4j Graph library to generate answers.

Brian concluded by framing these projects as a multi-faceted approach to making graph analytics more accessible, while also highlighting shared open challenges. These include the need for better benchmarks, improving the global planning, and handling billion-scale graphs. %with tools with large outputs.

\subsection{Chat2Graph: Graph Agentic System}

\medskip
\medskip

Dr. Heng Lin from Ant Group gave the third talk on their Chat2Graph.
Heng began by contextualizing the immense scale of Ant Group's business, which is inherently interconnected and operates on graphs with up to $10^{13}$ edges, posing significant challenges in performance, usability, and scale. This drove their decade-long development of the TuGraph graph database.

Heng continued by noting that methods like Text2GQL and GraphRAG \cite{edge2024local} are valuable.
For example, GraphRAG is used in Alipay's voice assistant to improve recall over traditional vector methods.
However, \textit{these approaches are insufficient for the complex analytical tasks in practice}.
To address this, his team is developing Chat2Graph \cite{chat2graph}, an agentic system that is ``graph-native'', using graphs not just as a data source, but as a core structure for its own reasoning and operations.

The talk detailed three primary components of the Chat2Graph architecture. The ``{\it Planning Graph}'' models complex task dependencies for strategic planning and rescheduling, moving beyond simple linear to-do lists. The ``{\it Toolkit Graph}'' organizes and maps available tools to specific actions, enabling intelligent tool discovery and recommendation. Finally, the ``{\it Memory Graph}'' implements a hierarchical memory system that organizes the agent's experiences into layers, escalating raw data from History up to strategic Insights to create a more interpretable and efficient long-term memory.

Heng concluded by presenting Chat2Graph, a graph native agentic system, which is a comprehensive, open-source ecosystem designed to handle complex, real-world graph-centric problems.
%The system's vision is to leverage graphs as a foundational element for building more powerful and intelligent agents.

%% file: 4-RESEARCH.tex
\section{Research Papers}

\medskip
\medskip
The peer-reviewed research papers presented in this workshop can be broadly classified into two categories, which are discussed in the following two subsections.

\textbf{LLMs for Graph Construction \& Reasoning.}
Constructing and maintaining high-quality graphs is a significant challenge due to the high cost of manual curation and extracting structured information.
Furthermore, validating graph consistency and interpreting the results often requires specialized expertise.
Papers in this category explore how the natural language understanding and generation capabilities of LLMs can be leveraged to address these challenges.
Colombo et al.~\cite{colombo2150llm} present a pipeline using fine-tuned LLMs to construct a knowledge graph of U.S. public laws from unstructured sources, including low-quality PDFs.
Ji et al.~\cite{ji2150llm} introduce a framework to evaluate how well twelve representative LLMs perform taxonomic reasoning, using graph-based test cases to assess their capabilities systematically.
Publio et al.~\cite{publio2150xpshacl} propose an explainable SHACL validation system that uses RAG and LLMs to produce human-readable, multi-language explanations for constraint violations in KGs. 
Mihindukulasooriya et al. \cite{mihindukulasooriya2025automatic} empirically study automatic prompt optimization techniques like DSPy~\cite{khattab2024dspy} and APE~\cite{zhou2022large} for triple extraction, showing these methods generate high-quality prompts that improve KG construction performance.

\textbf{Graphs for LLM Enhancement \& Retrieval.}
To address common LLM limitations like hallucinations and a lack of domain-specific knowledge, the papers in this category investigate how graphs can serve as a structured external knowledge source.
This approach aims to ground LLMs, enhancing their performance, scalability, and reasoning within RAG and agentic systems.
Zhang et al.~\cite{zhang2150scalable} propose using Locality-Sensitive Hashing to address the scalability challenges of tree-based Graph-RAG, improving the efficiency of graph construction and query retrieval on large datasets.
Schneider et al.~\cite{schneider2150graph} address the limitations of LLMs in spatial reasoning and envision a future where graph-enhanced reasoning is integrated into search engines to answer complex spatial questions. Zhou et al.~\cite{zhou2150towards} propose a conceptual framework for intelligent agent systems consisting of four key modules: planning, execution, knowledge, and tools, outlining future research directions of agentic systems.

%% file: 5-PANEL.tex
\section{Panel}

\medskip
\medskip
The panel focused on the theme of ``LLMs vs. Graphs: Supercharge or Supersede Graph Data Management, Mining, and Learning?''
The panelists were Reynold Cheng (University of Hong Kong), Philip S. Yu (University of Illinois, Chicago), Angela Bonifati (Lyon 1 University), Heng Lin (Ant Group), and Yan Da (Indiana University  Bloomington). The discussion, structured around five core themes, explored the evolving relationship between LLMs and graph technologies.

The panel first addressed the {\it synergy between LLMs and graph computation, debating whether they are in competition or collaboration}. Panelists agreed that LLMs should not supersede core graph database functionalities but rather serve as natural language interfaces (e.g., for Text-to-Graph-Query) and tools for integrating unstructured text in GraphRAG scenarios.
%Prof. Reynold Cheng noted that this area requires more research and benchmarks.
A key challenge identified was building feedback loops where the graph's factual knowledge can verify and improve LLM accuracy.

Discussing the {\it evolution of graph data management for LLM workloads}, panelists noted that classic challenges like query optimization and indexing remain critical, where LLMs might aid in explainability. They also explored novel uses of graphs to support AI, such as structuring agentic systems with a ``planning graph'' or generating high-quality training data.

Regarding the \textit{future of education and training}, panelists agreed that computer science is at a ``crossroad'', requiring a curricular shift towards interdisciplinary project-based learning. The consensus was that rather than forbidding LLMs, students should be taught to use them critically, focusing on skills like debugging AI-generated code over repetitive tasks. Stronger industry-academia partnerships were also emphasized as vital.

Finally, the discussion highlighted the \textit{foundational challenge of knowledge graph construction}, a difficult conceptual modeling problem where the database community's expertise in schema design is indispensable and where LLMs currently fall short. The panel concluded by reinforcing the vision of a synergistic future, with many open research questions remaining in system design, data management, and education.

% A significant portion of the discussion was dedicated to {\it the future of education and training}.
% The panelists agreed that computer science education is at a "crossroad" and that curricula require fundamental rethinking in the age of AI. There was a strong sentiment for moving towards more interdisciplinary, project-based learning that teaches students how to integrate knowledge from different domains. Rather than forbidding the use of LLMs, educators should teach students to use them critically—focusing on skills like debugging AI-generated code and high-level system design, rather than on repetitive programming tasks that can be automated. Stronger industry-academia partnerships, through open-source projects in classrooms and jointly developed curricula, were seen as vital to preparing the next generation of talent.

% Finally, the discussion touched upon the {\it foundational challenge of knowledge graph construction from unstructured data.} 
% The panelists emphasized that this remains a difficult conceptual modeling problem where the database community's deep-rooted knowledge in schema design and data modeling is indispensable—tasks that LLMs currently cannot perform reliably on their own. The panel concluded by reinforcing the vision of a synergistic future, with significant open research questions in system design, data management, and education to fully realize the potential of combining LLMs and graphs.

%% file: 6-Future.tex
\section{Future Directions}

\medskip
\medskip
We identify several active directions in the emerging area of LLM+Graph, listed as follow:

$\bullet$ \textbf{Foundational Models and Advanced Reasoning on Graphs}. Progress requires general-purpose models with a deeper grasp of graph structures, moving beyond retrieval-based approaches toward multi-hop reasoning, structural abstraction, and stronger generalization across diverse graph tasks.

$\bullet$ \textbf{Scalable and Unified Data Systems.} Building robust infrastructures that integrate LLMs with graph data remains a major challenge. Future systems must support hybrid queries over large graphs and vector embeddings, while adapting to the dynamic nature of real-world data. 

$\bullet$ \textbf{Trustworthy and Interactive AI on Graphs.} Adoption hinges on reliability and transparency. Key directions include automated data cleaning, consistency enforcement, explainable outputs to foster user trust, and human-in-the-loop frameworks for tasks requiring expert oversight